\documentclass[10pt,conference]{IEEEtran}
\usepackage{cite,graphicx,psfrag,amsmath,amssymb,amsfonts,verbatim}

\newcommand{\sr}{\stackrel}

\newcommand{\rar}{\rightarrow}

\newcommand{\tri}{\sr{\triangle}{=}}

\newcommand{\be}{\begin{equation}}
\newcommand{\ee}{\end{equation}}
\newcommand{\bea}{\begin{eqnarray}}
\newcommand{\eea}{\end{eqnarray}}
\newcommand{\bes}{\begin{eqnarray*}}
\newcommand{\ees}{\end{eqnarray*}}
\newcommand{\bi}{\begin{itemize}}
\newcommand{\ei}{\end{itemize}}
\newcommand{\ben}{\begin{enumerate}}
\newcommand{\een}{\end{enumerate}}
\newcommand{\bp}{\begin{problem}}
\newcommand{\ep}{\end{problem}}
\newcommand{\hso}{\hspace{.1in}}
\newcommand{\hst}{\hspace{.2in}}
\newcommand{\hsf}{\hspace{.5in}}

\begin{document}
\title{Minimum Redundancy Coding for Uncertain Sources}

\author{\authorblockN{Michael~B.~Baer}
\authorblockA{Vista Research\\
Monterey, CA,  USA\\
Email: calbear\hspace{1sp}@\hspace{1sp}ieee.org}
\and
\authorblockN{Farzad Rezaei}
\authorblockA{Email: frezaei\hspace{1sp}@\hspace{1sp}alumni.uottawa.ca} \and
\authorblockN{Charalambos D. Charalambous}
\authorblockA{Dept.\ of Electrical and Computer Eng.\\
University of Cyprus\\
Nicosia, Cyprus\\
Email: chadcha\hspace{1sp}@\hspace{1sp}ucy.ac.cy}\\
 }
\maketitle

\begin{abstract}
Consider the set of source distributions within a fixed maximum relative entropy with respect to a given nominal distribution.  Lossless source coding over this relative entropy ball can be approached in more than one way.  A problem previously considered is finding a minimax average length source code.  The minimizing players are the codeword lengths --- real numbers for arithmetic codes, integers for prefix codes --- while the maximizing players are the uncertain source distributions.  Another traditional minimizing objective is the first one considered here, maximum (average) redundancy.  This problem reduces to an extension of an exponential Huffman objective treated in the literature but heretofore without direct practical application.  In addition to these, this paper examines the related problem of maximal minimax pointwise redundancy and the problem considered by Gawrychowski and Gagie, which, for a sufficiently small relative entropy ball, is equivalent to minimax redundancy.  One can consider both Shannon-like coding based on optimal real number (``ideal'') codeword lengths and a Huffman-like optimal prefix coding.
\end{abstract}


\section{Introduction}\label{intro}

\subsection{Preliminaries}

The well-known problem of finding a uniquely decodable code with minimum average codeword length over a memoryless source gives rise to the optimal Huffman code and the near-optimal Shannon code.  The derivation of the latter code assures that average redundancy of neither code exceeds~$1$.  However, suppose the true distribution of the source is unknown and the code is designed solely with respect to a nominal distribution, for example one derived from access to limited empirical data.  In this case the relative entropy between the nominal distribution and the true distribution appears in the redundancy bounds of average codeword length; indeed, the lower bound is the sum of the entropy and the divergence\cite[Theorem 5.4.3]{cover-thomas06}.  Consequently, under such uncertainty, the nominally optimal codes will be robust neither in average length nor in redundancy.

Suppose the nominal or approximate distribution of a source is $\mu$, while the unknown or true probability distribution of the source is any $\nu$ which is absolutely continuous with respect to $\mu$ and satisfies a relative entropy constraint.  This constraint is ${\mathbb D}(\nu\|\mu) \leq R$, where $R$ is a known, positive entropy value (in nats) and
\bes
{\mathbb D}(\nu\|\mu) \tri \sum_i \nu_i \log \frac{\nu_i}{\mu_i} \label{D}
\ees
where the sum is over all events and $\log$ is the natural logarithm.

Source coding problems are often idealized so that codeword lengths need not be integer; this is useful for arithmetic codes, for calculating limits for arbitrarily large blocks, and for finding approximate solutions, and, by extension, performance bounds.  Such approximate solutions often require less computation and/or memory to compute than optimal solutions, and these robust Shannon-like codes will be our intended application of the relaxation of $\ell_i \in \mbox{\bf Z}_+$ to $\ell_i \in \mbox{\bf R}$ (with no further mention of arithmetic codes and other related applications).

We can assume the $M$-member input alphabet is, without loss of generality, equal to ${\mathcal X} \tri \{1,2, \ldots ,M\}$.  Denote the class of all uniquely decodable codes defined on ${\mathcal X}$ by ${\cal C}({\mathcal X})$ or ${\cal C}_M$.  We use codes and codeword length ensembles interchangeably, since one can easily be derived from the other; thus valid solutions can be said to satisfy $\ell \in {\cal C}({\mathcal X})$ for random vector~$\ell$.  An ensemble of lengths corresponds to a uniquely decodable code if and only if all values are natural numbers and satisfy the Kraft inequality $\sum D^{-\ell_i} \leq 1$.  Denoting the set of real vectors which satisfy the Kraft inequality as ${\cal K}({\mathcal X})$, we can mathematically restate the previous sentence as ${\cal C}({\mathcal X}) = {\cal K}({\mathcal X}) \cap \mbox{\bf Z}_+^M$ where $\mbox{\bf Z}_+^M$ is the set of all $M$-vectors of positive integers.

\subsection{Objectives}
\label{objectives}

There are several possible approaches to such a source coding problem.

{\it (Average) Minimax Length Approach}: This approach, explored in
\cite{charalambous-rezaei05}, is concerned with the minimax average 
length formulation which, with a relative entropy constraint, is
\be
\inf_{\ell^\ddagger \in {\cal K}({\mathcal X})} \sup_{\{\nu; {\mathbb D}(\nu\|\mu) \leq R\}}{\mathbb E}_{\nu}(\ell^\ddagger) \label{shanopt}
\ee
for \textit{robust Shannon code} $\ell_i^\S = \lceil \ell_i^\ddagger \rceil$ and
\be
\inf_{\ell^* \in {\cal C}({\mathcal X})} \sup_{\{\nu; {\mathbb D}(\nu\|\mu) \leq R\}}{\mathbb E}_{\nu}(\ell^*) \label{expopt}
\ee
for \textit{robust Huffman code} $\ell^*$, where ${\mathbb E}_\nu$ denotes expectation with respect to the distribution~$\nu$.  The main objective of the minimax formulation is to encode the output of uncertain sources using the worst case distribution $\nu^*$, i.e., the distribution which maximizes the average length as a function of the nominal distribution $\mu$.  By judging the coding of the uncertain source according to $\nu^*$, the resulting code will be robust in the sense of average length over the set of uncertain sources which satisfy the relative entropy constraint.

{\it (Average) Minimax Redundancy Approach}: This approach, explored in Section~\ref{Redundancy}, finds codes robust in the sense of average redundancy.  The minimax redundancy (or average minimax redundancy) formulation is
\be
\inf_{\ell^{\textrm{\tiny{AL}}} \in {\cal C}({\mathcal X})}
\sup_{\{\nu; {\mathbb D}(\nu\|\mu) \leq R \}} {\mathbb E}_{\nu}(\ell^{\textrm{\scriptsize{AL}}}) - {\mathbb H}_D(\nu) 
\label{redopt}
\ee for the Huffman-style problem, where ${\mathbb H}_D(\nu)$ is the entropy of the source with distribution $\nu$ in terms of compressed $D$-ary symbols.  The Shannon counterpart is trivially the Shannon code $\ell_i^{\textrm{\scriptsize{shannon}}} = \lceil - \log_D \mu_i \rceil$ for~$\mu$, since \textit{ideal codeword length} $\ell_i^\dagger \tri - \log_D \mu_i$ has an expectation equal to entropy plus relative entropy\cite[Section~13.1]{cover-thomas06}.  Because of this, in the idealized problem --- no matter what the domain for $\nu$ --- it is sufficient to find the smallest relative entropy ball that contains the set and use the optimal code around which the entropy ball is centered.  Thus, a divergence entropy ball is a ``natural set''\cite[Section 13.1]{cover-thomas06}, also seen in past treatments of optimal codes and related problems\cite{longo-galasso82,Dunham92,charalambous-rezaei07}.  This is closely related to the method of types, in which relative entropy of observed distribution $\nu$ relative to actual distribution $\mu$ is closely related to the probably of a given type\cite{csiszar-korner81,Csiszar98}.

{\it Gawrychowski-Gagie Approach}\cite{gawrychowski-gagie09}: This approach, like the average redundancy, is based on the premise of minimizing the average amount by which codeword lengths exceed what they ``should.''  In this case, rather than using the amount average codeword length exceeds the entropy of $\nu$ --- the ideal average length {\it with} knowledge of $\nu$ --- we use the amount it exceeds the entropy of $\nu$ plus the relative entropy of $\nu$ with respect to $\mu$ --- the ideal average length {\it without} knowledge of~$\nu$:
\be
\inf_{\ell^{\textrm{\tiny{GG}}} \in {\cal C}({\mathcal X})}
\sup_{\{\nu; {\mathbb D}(\nu\|\mu) \leq R \}} {\mathbb E}_{\nu}(\ell^{\textrm{\scriptsize{GG}}}) - {\mathbb H}_D(\nu) - {\mathbb D}_D(\nu\|\mu)
\label{ggopt}
\ee for the Huffman-style problem, where ${\mathbb D}_D(\nu\|\mu) \tri
(\log D)^{-1}{\mathbb D}(\nu\|\mu)$, the relative entropy using
base~$D$.  Again, the Shannon version is trivial.  However, unlike the
other utilities discussed, here the optimized value includes a term
from the nominal distribution~$\mu$; we look at this utility in association 
with the previous one in Section~\ref{Redundancy}.

{\it Maximal Minimax Redundancy Approach}: This approach, considered in Section~\ref{pointwise}, is concerned with the maximal minimax (pointwise) redundancy formulation.  This type of redundancy takes the maximum difference between $\ell_i$ and $\ell_i^\dagger = - \log_D \mu_i$ rather than the expected difference.  This results in
\be
\inf_{\ell^{\textrm{\tiny{ML}}} \in {\cal C}({\mathcal X})}
\sup_{\{\nu; {\mathbb D}(\nu\|\mu) \leq R \}} \max_{k \in [1,M]} (\ell_k^{\textrm{\scriptsize{ML}}} + \log_D \nu_k)
\label{mmhopt}
\ee
in the Huffman case and
\bea
\inf_{\ell^{\textrm{\tiny{MI}}} \in {\cal K}({\mathcal X})}
\sup_{\{\nu; {\mathbb D}(\nu\|\mu) \leq R \}} \max_{k \in [1,M]} (\ell_k^{\textrm{\scriptsize{MI}}} + \log_D \nu_k)
\label{mmsopt}
\eea
in the real case from which a Shannon-like code can be derived.  We investigate (\ref{redopt})-(\ref{mmsopt}) here.

\subsection{Literature Review}

There is a significant literature dealing with source coding for
an unknown source when either the empirical distribution of the source
is available or the uncertainty is modeled through certain unknown
parameters \cite{lddavisson73,davisson-leongarcia80,rissanen84}. A
good survey of the literature is found in
\cite{barron98}. Uncertainty modeling using relative entropy has been
considered in \cite{barron92}, in which the problem of coding for only
one unknown source is addressed.  However, unlike \cite{barron98} and
\cite{barron92}, here we are dealing with the problem of source coding
for a class containing many sources which satisfy the relative entropy
constraint.  Our modeling assumes a knowledge of the nominal
distribution and the uncertainty radius~$R$. Universal modeling
using relative entropy has been discussed in~\cite{Dunham92}, where
the tightest upper bound for the relative entropy between empirical
distributions (of available training sequences), and a nominal
distribution is found. The nominal distribution is itself computed as
part of a search algorithm.  Longo and Galasso considered relative 
entropy balls in which the same code was optimal for all probability 
mass functions\cite{longo-galasso82}.
Universal utility (\ref{ggopt}) is taken over the entire simplex in 
\cite{gawrychowski-gagie09}, while (\ref{mmhopt}) is considered for
arbitrary sets in \cite{Szpank2004} and \cite{Szpank2004a}.

Finally, we point out that the current minimax source coding
formulation may be generalized to applications in which the nominal
distribution is parameterized as in \cite{barron98}.  In this case, the
methods found in \cite{barron98} which employ maximum likelihood
techniques can be invoked to estimate the parameters of the nominal
distribution.  Additional generalizations may include situations in
which the source is described by ergodic finite-state Markov chains.

\section{Minimax Redundancy Coding}
\label{Redundancy}
Let ${\cal M}({\mathcal X})$ denote the set of all probability mass functions defined on~${\mathcal X}$.  Nominal probability distribution $\mu \in {\cal M}({\mathcal X})$ is known; we assume that any (unknown) true probability distribution $\nu$ is absolutely continuous with respect to $\mu$ in ${\cal M}({\mathcal X})$, i.e., if $\mu_i=0$ for any $i \in {\mathcal X}$, then $\nu_i=0$.  In addition, we know non-negative $R$ defining the domain of possible solutions, ${\cal M}_R = \{ \nu \in {\cal M}({\mathcal X}); {\mathbb D}(\nu\|\mu) \leq R\}$.  Given nominal $\mu \in {\cal M}({\mathcal X})$ and non-negative $R$, the problem is to find the codeword lengths $\{\ell_j^*\}$ of a uniquely decodable code and a probability measure $\nu^* \in {\cal M}_R$ which solve the following generalized minimax source coding problem:
\bea
\left\{
\begin{array}{l}
 J_f(\ell^*,\nu^*)=  \inf_{(\ell_1,\ldots,\ell_M)} \sup_{\nu \in {\cal M}({\mathcal X})}
f(\ell,\nu) \\
 \mbox{Subject to}\hso  {\mathbb D}(\nu\|\mu) \leq R, \hst \sum_{i=1}^{M} D^{-\ell_i} \leq 1
\end{array}
\right. \label{general}
\eea where $f$ is the utility in question.  
Values $\{\ell_i\}$ are the lengths of the 
$D$-ary codewords.  Encoding based on the worst case measure $\nu^* 
\in {\cal M}_R$ results in average codeword length being less sensitive to 
different source distributions within set ${\cal M}_R$.

Recall
\be
\inf_{\ell^{\textrm{\tiny{AL}}} \in {\cal C}({\mathcal X})}
\sup_{\{\nu; {\mathbb D}(\nu\|\mu) \leq R \}} \left({\mathbb E}_{\nu}(\ell^{\textrm{\scriptsize{AL}}}) - {\mathbb H}_D(\nu)\right) \tag{\ref{redopt}}
\ee
and the closely related
\be
\inf_{\ell^{\textrm{\tiny{GG}}} \in {\cal C}({\mathcal X})}
\sup_{\{\nu; {\mathbb D}(\nu\|\mu) \leq R \}} \left({\mathbb E}_{\nu}(\ell^{\textrm{\scriptsize{GG}}}) - {\mathbb H}_D(\nu) - {\mathbb D}_D(\nu\|\mu)\right)
\tag{\ref{ggopt}}
\ee
two specific instances of (\ref{general}).
The former, minimax redundancy (or average minimax redundancy), is one of the most widely used measures of performance for designing codes when the source is subject to  uncertainty (e.g., \cite{lddavisson73}, \cite{davisson-leongarcia80}, and \cite{davisson-mceliece-Pursely-Wallace81}).  As it turns out, the minimax solution to this problem leads to encoding with an exponential pay-off similar, but not identical, to (2) in \cite{charalambous-rezaei05}.  This encoding, where it applies, also solves (\ref{ggopt}).

Recall $\{\ell_1,\ldots,\ell_M \}$ denotes codeword lengths for the source symbols $\{1,\ldots,M\}$. Assume $\nu \in {\cal M}_R$, which implies ${\mathbb D}(\nu \|\mu) \leq R$. Let $r(\ell, \nu)$ denote the redundancy of the code. Formulate the problem of average minimax redundancy as:
\bes
\lefteqn{
\inf_{(\ell_1,\ldots,\ell_M)} \sup_{\nu \in {\cal M}_R} \left( {\mathbb E}_{\nu}(\ell)-{\mathbb H}_D(\nu)\right)}\\
\quad &=&\inf_{(\ell_1,\ldots,\ell_M)} \sup_{\nu \in {\cal M}_R} r(\ell, \nu).
\ees
Average redundancy can be written as follows:
\be
\begin{array}{rcl}
r(\ell, \nu)&=& \frac{1}{\log D}{\mathbb D}(\nu\|\theta) \\
&=& \frac{1}{\log D} \left({\mathbb D}(\nu \|\mu) + \sum_{i=1}^{M} \nu_i \log \left( \frac{\mu_i}{\theta_i}\right) \right)
\end{array}
\label{f111_Ch1}
\ee
where $\theta_i \tri D^{-\ell_i}$ for all $i \in \{1,\ldots,M\}$. Now consider the following probability distribution.
\bea
\nu^{\circ}_i(\beta) \tri \frac{\left( \frac{\mu_i}{\theta_i}\right)^{\beta} \mu_i}{ \sum_{k=1}^M \left( \frac{\mu_k}{\theta_k}\right)^{\beta} \mu_k}  \hsf \beta > 0 \label{nuo}
\eea
The relative entropy between $\nu$ and $\nu^{\circ}(\beta)$ can be found as follows.
\be
\begin{array}{rcl}
{\mathbb D}(\nu \| \nu^{\circ}(\beta)) &=& {\mathbb D}(\nu \| \mu) + \log \left(\sum_{k=1}^M \left( \frac{\mu_k}{\theta_k}\right)^{\beta} \mu_k\right) \\
&& {} - \beta \sum_{i=1}^{M} \nu_i \log \left( \frac{\mu_i}{\theta_i}\right)
\end{array}
\label{f110_Ch1}
\ee
Now substitute $\sum_{i=1}^{M} \nu_i \log \frac{\mu_i}{\theta_i}$ from (\ref{f110_Ch1}) in (\ref{f111_Ch1}). Then
\be
\begin{array}{rcl}
r(\ell, \nu) &=& \frac{1}{\log D} \left( \frac{\beta + 1}{\beta} {\mathbb D}(\nu \| \mu) - \frac{1}{\beta} {\mathbb D}(\nu \| \nu^{\circ}(\beta))\right. \\
&&\left.{} + \frac{1}{\beta} \log \left(\sum_{k=1}^M \left( \frac{\mu_k}{\theta_k}\right)^{\beta} \mu_k\right)\right) \label{f112}
\end{array}
\ee
The supremum of (\ref{f112}) is attained at $\nu = \nu^{\circ}(\beta)$, where $\beta$ is the value such that ${\mathbb D}(\nu^{\circ}(\beta) \| \mu) = R$, for those problems in which such a $\beta>0$ exists.  This maximizes $\beta$ within the relative entropy constraint, thus maximizing the first term, bringing the second term to zero, and maximizing the third term (due to Lyapanov's inequality for moments, an application of H\"{o}lder's inequality, e.g., \cite[p.~27]{HLP} or \cite[p.~54]{mitrinovic}).  Hence,
\bea
\sup_{\nu \in {\cal M}_R} r(\ell, \nu) = r(\ell, \nu^{\circ}(\beta))  \label{f113}
\eea
Therefore, minimizing the worst case redundancy subject to codeword lengths leads to the following subproblem.
\be
\begin{array}{l}
\inf_{(\ell_1,\ldots,\ell_M)} \frac{1}{\beta} \log_D \sum_{k=1}^M \left(
\frac{\mu_k}{\theta_k}\right)^{\beta} \mu_k \\
{} = \inf_{(\ell_1,\ldots,\ell_M)} \frac{1}{\beta} \log_D \sum_{k=1}^M e^{\beta (\log D) (\ell_k - \ell^{\prime}_k)}\mu_k
\end{array} 
\label{nath}
\ee
where $\ell^{\prime}_k = \log_D \frac{1}{\mu_k}$. This is a special case
of the general problem first posed in \cite{nath75} and also
considered in \cite{parker80} and \cite{Baer2006}.  Of these last two,
the former gives an optimal solution, while the latter notes that this 
solution is closely related to Shannon coding; for (\ref{f113}), a 
conventional Shannon code will not exceed the optimal code by more than 
one output symbol per input symbol.

The optimal solution is obtained using the exponential Huffman 
coding as described in \cite{hu-kleitman-tamaki79, parker80, humblet81,
charalambous-rezaei05} for the following cost function:
\bes \inf_{(\ell_1,\ldots,\ell_M)} \sum_{k=1}^M e^{s\ell_k} \xi_k
\ees where $s = \beta \log D$ and $\xi$ is a probability
distribution given by \bes \xi_k = \frac{\mu_k^{\beta +
1}}{\sum_{i=1}^{M} \mu_i^{\beta + 1}} \ees
This solution is linear time given sorted $\mu$.  The $\beta$ for
which ${\mathbb D}(\nu^{\circ}(\beta) \| \mu) = R$ can be found using
root-finding methods, such as those described for (\ref{expopt}) in
\cite{charalambous-rezaei05}.

As to the matter of whether such a $\beta$ exists, consider $\beta
\rar \infty$.  (Taking $\beta \rar 0$ corresponds to $R=0$.)  As shown
in \cite{Baer2006}, there exists a $\beta'$ such that, for all $\beta
\geq \beta'$, (\ref{nath}) is solved by the optimal codeword solution
to the minimax pointwise redundancy problem, the equation itself
becoming $\inf_\ell \max_k (\ell_k-\ell_k^{\prime})$.  Taking the
limit of (\ref{nuo}), we find the following maximal probability mass
function: \bes \nu^{\infty}_i = \lim_{\beta \rar \infty}
\nu^{\circ}_i(\beta) = \frac{\mu_i \mbox{\bf 1}_i^*}{\sum_{k=1}^M
  \mu_k \mbox{\bf 1}_k^*} \ees where $\mbox{\bf 1}_i^*$ is $1$ if
$\frac{\mu_i}{\theta_i} = \max_k \frac{\mu_k}{\theta_k}$ and $0$
otherwise.  (For example, if this ratio has a unique maximum with
index $k$, then $\nu^{\infty}$ is $1$ for $k$ and $0$ for all other
values.)  This results in \bea {\mathbb D}(\nu^{\infty} \| \mu) =
-\log \sum_{k=1}^M \mu_k \mbox{\bf 1}_k^*
\label{solvemm}
\eea
$\nu^{\infty}$ being the maximizer and the 
two-variable solution of \cite{Baer2006} the minimizer of the minimax problem when $R$ is ${\mathbb D}(\nu^{\infty} \| \mu)$.  If ${\mathbb D}(\nu^{\infty}
\| \mu)$ exceeds $R$, then the desired $\beta$ exists due to continuity;
otherwise, it does not.  In the case that it does not, the simplex
boundaries ($\nu_i^{\circ}(\beta) \geq 0$) come into play and the optimization no
longer conforms to our model.

Now recall the utility of Gawrychowski and
Gagie\cite{gawrychowski-gagie09}, shown at the beginning of this
section in (\ref{ggopt}), which can also be expressed as $r(\ell, \nu)
- (\log D)^{-1} {\mathbb D}(\nu \| \mu)$.  Note that the above
analysis for $\beta$ --- for values where it holds --- also holds
for this, since the three terms in (\ref{f112}) are identical; only
the multiplicative factors are different, and these do not affect the
optimization.  The analysis in the limit also holds.  In this case, 
the optimal minimax solution in the limit is the optimal code for 
all $R \geq -\log \min_i \mu_i$, not just $R = -\log \min_i \mu_i$.
This is a consequence of the minimax solution over the simplex, $R
\rightarrow \infty$, being the same\cite{gawrychowski-gagie09} (and
thus the solution for any superset of the aforementioned limit ball).
Note that while the two-variable solution is the limit, any minimax
pointwise solution --- i.e., any of possibly multiple codes minimizing
maximum redundancy, as is, e.g., \cite{Baer2006} --- suffices for
optimization.

\section{Maximal Minimax Redundancy Coding}\label{pointwise}
Recall
\be
\inf_{(\ell_1,\ldots,\ell_M)} \sup_{\{\nu; {\mathbb D}(\nu\|\mu) \leq R \}} 
\max_{k} (\ell_k + \log_D \nu_k)
\label{mm1}
\ee
as restricted in (\ref{mmhopt}) and~(\ref{mmsopt}).  This problem, 
closely related to that of average minimax redundancy,
is {\it maximal} minimax ({\it pointwise}) redundancy, another
robust coding measure considered in \cite{Szpank2004} based on earlier
work, e.g., \cite{shtarkov87,barron98}.  Pointwise
redundancy of item $k$ for a code with lengths $\ell$ given known 
probability mass function $\nu$ is equal to $\ell_k + \log_D \nu_k$, which is
the difference between codeword length and {\it self-information}.
Self-information, equal to $-\log_D \nu_k$, is the optimal codeword
length of the coding problem where $\nu$ is known and there is no
integer restriction.  Therefore minimax redundancy is average minimax
pointwise redundancy, or, over an arbitrary set of probabilities
${\cal N} \subseteq {\cal M}({\mathcal X})$ (not necessarily a relative
entropy ball),
\bes
\inf_{(\ell_1,\ldots,\ell_M)} \sup_{\nu \in {\cal N}} \sum_{k=1}^M 
(\ell_k + \log_D \nu_k) \nu_k
\ees
while maximal minimax pointwise redundancy is
\bes
\inf_{(\ell_1,\ldots,\ell_M)} \sup_{\nu \in {\cal N}}
\max_{k} (\ell_k + \log_D \nu_k).
\ees
Since this is equal to 
\bes 
\inf_{(\ell_1,\ldots,\ell_M)} \max_{k \in [1,M]} (\ell_k + \log_D 
\sup_{\nu \in {\cal N}} \nu_k)
\ees
we can find the supremums and then calculate the solution 
to the reduced problem appropriately, whether the robust Shannon or robust 
Huffman solution is desired.  
The supremums form what is called a 
{\it normalized maximum-likelihood (NML) distribution}, $\underline{\pi}$,
which is the normalized version of
\bes
\pi_k = \sup_{\nu \in {\cal N}} \nu_k
\ees
that is,
\bes
\underline{\pi}_k = \frac{\sup_{\nu \in {\cal N}} \nu_k}{\sum_{i=1}^M \sup_{\nu \in {\cal N}} \nu_i}.
\ees

If ${\cal N}$ is the set of probability mass functions within a certain total
variation $T$ of a known $\mu$, this is trivial to compute as $\pi_k = 
\min(1,\mu_k + T/2)$, and the solution can be found based on this.  In 
the case considered here, that of a relative entropy ball, ${\cal N}$ is ${\cal M}_R$ 
(that is, $\{\nu ; {\mathbb D}(\nu \| \mu) \leq R \}$).
The normalization, resulting in a constant difference in the minimized utility,
is optional for the purpose of building a robust Huffman code, since
the Huffman procedure --- efficiently done in \cite{Baer2006} for sorted 
probabilities and \cite{gawrychowski-gagie09} for unsorted probabilities
--- is scale-invariant.  The robust Shannon solution is based
on the optimal solution with the integer constraint removed; this optimal
solution is $-\log_M \underline{\pi}_k$ and thus the robust Shannon code
is $\ell^{\textrm{\scriptsize{MS}}} = \lceil -\log_M \underline{\pi}_k \rceil$.

Because finding the optimal solution follows from finding the maximum likelihood distribution, the only unaddressed issue is the following: Given probability mass function $\mu$, index $k$, and relative entropy $R$, find (non-normalized) vector $\pi^{(k)}$ which maximizes $\pi_k^{(k)}$ within the constraint ${\mathbb D}(\pi^{(k)} \| \mu) \leq R$, so that the non-normalized maximum likely distribution has $\pi_k = \pi_k^{(k)}$.

First let us denote the deterministic distribution with all its weight on 
item $k$ as $\omega^{(k)}$, the probability distribution such that
$\omega_k^{(k)} = 1$ (i.e., for $j \neq k$, $\omega_j^{(k)} = 0$).  For each $k$, 
we should first check whether ${\mathbb D}(\omega_k \| \mu) \leq R$.  If so,
clearly $\pi_k^{(k)} = 1$.  This is the case if
\bes
{\mathbb D}(\omega_k \| \mu) &=& 1 \cdot \log \frac{1}{\mu_k} + \lim_{x \rar 0} 
\sum_{i \neq k} x \log \frac{x}{\mu_i} \\
&=& -\log \mu_k \leq R
\ees
which occurs if and only if $\mu_k \geq e^{-R}$.

For those values, if any, not satisfying this, we take a Lagrangian approach to this constrained minimization, reducing this problem to several problems each of a single dimension with roots that can be easily found.  Specifically, if $s$ and $\lambda$ are the Lagrangian multipliers, and $I_{i,k}$ is a function that is $1$ if $i=k$ and $0$ otherwise, then to maximize
\bes
L^{\lambda,s}(\pi^{(k)}) &=& \pi_k^{(k)} + s \left(\sum_{i=1}^M \pi_i^{(k)} \log \frac{\pi_i^{(k)}}{\mu_i} - R \right) \\
&&{} + \lambda \left(\sum_{i=1}^M \pi_i^{(k)} - 1 \right)
\ees
we require that
\bes
\frac{\partial L^{\lambda,s}(\pi^{(k)})}{\partial \pi_i^{(k)}} =
I_{i,k} + s(1+\log \frac{\pi_i^{(k)}}{\mu_i}) + \lambda = 0 
\ees
which means that $\pi_i^{(k)}$ is proportional to $\mu_i$ for all $i \neq k$.  
Thus, given probability mass function $\mu$, index $k$, and relative entropy 
$R$, the maximizing $\pi^{(k)}$ has $\pi_i^{(k)} = \rho_k \mu_i$ for some $\rho_k$ 
on all $i \neq k$ (and thus $\pi_k^{(k)} = 1 - \rho_k(1-\mu_k)$), so we actually 
only need to solve binary divergence
\bes
R = {\mathbb D}(\pi^{(k)} \| \mu) &=& \pi_k^{(k)} \log \frac{\pi_k^{(k)}}{\mu_k} + 
\sum_{i \neq k}  \pi_i^{(k)} \log \rho_k \\
&=& \pi_k^{(k)} \log \frac{\pi_k^{(k)}}{\mu_k} + 
(1-\pi_k^{(k)}) \log 
\frac{1-\pi_k^{(k)}}{1-\mu_k} \\
&=& \mbox{\bf d}(\pi_k^{(k)}\|\mu_k) \\
\ees
for the larger of the two possible solutions, where
\bes
\mbox{\bf d}(p\|m) \tri p \log \frac{p}{m} + 
(1-p) \log \frac{1-p}{1-m}.
\ees
This solution lies in the range 
\bea
\pi_k^{(k)} \in \left(\mu_k,\mu_k + \sqrt{\frac{1}{2} R}\right] \cap (0,1) \label{solrange}
\eea
with the lower bound due to $\mbox{\bf d}(\mu_k\|\mu_k) = 0$ and the non-trivial
upper bound derived from Pinsker's inequality, due to 
\bes
\pi_k^{(k)}-\mu_k &=& \frac{1}{2}\left\|[\pi_k^{(k)}~1-\pi_k^{(k)}] - [\mu_k~1-\mu_k]\right\|_{\textrm{\scriptsize{tv}}} \\
&\leq& \frac{\sqrt{2}}{2} \mbox{\bf d}(\pi_k^{(k)}\|\mu_k)
\ees
For small $R$, a
closed-form approximation can also be obtained under the assumption
that $\pi_k^{(k)} \leq 2 \mu_k$.  Taking terms up to the second
order for the power series of the logarithm, we find
\bes
R &=& \pi_k^{(k)} \log \left(\frac{\pi_k^{(k)}}{\mu_k}\right)
+ (1-\pi_k^{(k)})\log \left(\frac{1-\pi_k^{(k)}}{1-\mu_k}\right) \\
&=& \frac{(\pi_k^{(k)}-\mu_k)^2}{2(1-\mu_k)^2\mu_k^2} \left(2 \mu_k - \pi_k^{(k)} - 3 \mu_k^2 + 2 \mu_k \pi_k^{(k)} \right) \\
&& {} + O\left({\left(\frac{\mu_k-\pi_k^{(k)}}{\mu_k}\right)}^3 \pi_k^{(k)}\right) \\
&& {} + O\left({\left(\frac{\pi_k^{(k)}-\mu_k}{1-\mu_k}\right)}^3 (1-\pi_k^{(k)})\right) \\
&=& \frac{(\pi_k^{(k)}-\mu_k)^2}{2(1-\mu_k)\mu_k} + O\left(\frac{(\pi_k^{(k)}-\mu_k)^3}{(1-\mu_k)^2\mu_k^2}\right)
\ees
where we use $\pi_k^{(k)} = O(\mu_k)$ to simplify the additive order term.
We can remove $\pi_k^{(k)}$ from this term using Pinsker's 
inequality, so that
\bes
\pi_k^{(k)} = \mu_k + \sqrt{2 R (1-\mu_k) \mu_k + O\left(\frac{{R}^{\frac{3}{2}}}{(1-\mu_k)\mu_k}\right)}
\ees
This can be used to find each of the $k$ (approximated) solutions 
or as a first guess in Newton's or Halley's method to find solutions with 
arbitrary precision.  Both these methods converge quickly (quadratic and 
cubic convergence, respectively), as the function is convex increasing and 
three-times continuous differentiable with finite non-zero derivatives and no other zeroes over the first two derivatives in the solution range (\ref{solrange}). 
The solution $\pi$ can then be used to construct the corresponding robust Shannon or robust Huffman (minimax pointwise redundancy) code.  Note that in this case, the robust Huffman code will have the property that no length is longer than that of the robust Shannon code; if it were, the utility minimized would be at least $1$, whereas the robust Shannon code shows that the minimum should be less than $1$.  Thus, even in a pointwise sense, the robust Huffman code is not inferior to the robust Shannon code.

\section*{Acknowledgment}

The work of C.D.~Charalambous has received funding from the European
Community's Seventh Framework Programme (FP7/2007-2013) under grant agreement no.~INFSO-ICT-223844.

\bibliographystyle{IEEEtran}

\begin{thebibliography}{1}

\bibitem{cover-thomas06}
T.~Cover and J.~Thomas, {\em Elements of Information Theory}, Second Edition.
\newblock Wiley-Interscience, 2006.

\bibitem{charalambous-rezaei05}
C.D.~Charalambous and F.~Rezaei, ``Robust Coding for Uncertain Sources: A Minimax Approach,'' {\em Proc., 2005 IEEE Int. Symp. on Information Theory}, Adelaide, SA, Australia, pp.~1539-1543, Sep.~2005.

\bibitem{longo-galasso82}
G.~Longo and G.~Galasso, ``An Application of Informational Divergence to {Huffman} Codes,'' {\em IEEE Transactions on Information Theory}, vol.~IT-28, pp.~36--43, Jan.~1982.

\bibitem{Dunham92}
C.C.~Lu and J.G.~Dunham, ``A Universal Model Based on Minimax Average Divergence,'' {\em IEEE Transactions on Information Theory}, vol.~IT-38, pp.~140--144, Jan.~1992.

\bibitem{charalambous-rezaei07}
C.D.~Charalambous and F.~Rezaei, ``Stochastic Uncertain Systems Subject to
  Relative Entropy Constraints: Induced Norms and Monotonicity Properties of
  Minimax Games,'' {\em IEEE Transactions on Automatic Control}, vol.~52,
  pp.~647--663, Apr.~2007.

\bibitem{csiszar-korner81}
I.~Csisz\'{a}r and J.~K\"{o}rner, {\em Information Theory: Coding Theorems for Discrete Memoryless Systems}.
\newblock Academic, 1981.

\bibitem{Csiszar98}
I.~Csisz\'{a}r, ``The Method of Types,'' {\em IEEE Transactions on Information Theory}, vol.~44,  pp.~2505--2523, Oct.~1998.

\bibitem{gawrychowski-gagie09}
P.~Gawrychowski and T.~Gagie, ``Minimax Trees in Linear Time with
  Applications,'' in {\em Lecture Notes in Computer Science, Proc., 20th
  International Workshop on Combinatorial Algorithms (IWOCA)}, Springer-Verlag,
  June 29, 2009.
\newblock arXiv:0812.2868v2.


\bibitem{lddavisson73}
L.~Davisson, ``Universal Noiseless Coding,'' {\em IEEE Transactions on
  Information Theory}, vol.~IT-19, pp.~783--795, Nov.~1973.


\bibitem{davisson-leongarcia80}
L.~Davisson and A.~Leon-Garcia, ``A Source Matching Approach to Finding Minimax
  Codes,'' {\em IEEE Transactions on Information Theory}, vol.~IT-26,
  pp.~166--174, Mar.~1980.

\bibitem{rissanen84}
J.~Rissanen, ``Universal Coding, Information, Prediction, and Estimation,'' {\em IEEE Transactions on Information Theory}, vol.~IT-30, pp.~629--636, July~1984.

\bibitem{barron98}
A.R.~Barron, J. Rissanen, B. Yu, ``The Minimum Description Length Principle in Coding and Modeling,'' {\em IEEE Transactions on Information Theory}, vol.~44,  pp.~2743--2760, Oct.~1998.

\bibitem{barron92}
A.R.~Barron, L.~Gy\"{o}rfi, E.C.~van der Meulen, ``Distribution Estimation Consistent in Total Variation and in two Types of Information Divergence,'' {\em IEEE Transactions on Information Theory}, vol.~38,  pp.~1437--1454, Sep.~1992.

\bibitem{Szpank2004}
M.~Drmota, W.~Szpankowski, ``Precise Minimax Redundancy and Regret,'' {\em IEEE Transactions on Information Theory}, vol.~50,  pp.~2686--2707, Nov.~2004.

\bibitem{Szpank2004a}
P.~Jacquet, W.~Szpankowski, ``Markov Types and Minimax Redundancy for Markov Sources,'' {\em IEEE Transactions on Information Theory}, vol.~50,  pp.~1393--1402, July~2004.

\bibitem{davisson-mceliece-Pursely-Wallace81}
L.~Davisson, R.~McEliece, M.~Pursely and M.~Wallace, ``Efficient Universal Noiseless Source Codes,'' {\em IEEE Transactions on Information Theory}, vol.~27,
  pp.~269--279, May~1981.


\bibitem{HLP}
G. H.~Hardy, J. E.~Littlewood and G.~Polya, {\em Inequalities}
\newblock Cambridge Univ.\ Press, 1934.

\bibitem{mitrinovic}
D. S.~Mitrinovic, {\em Analytic Inequalities}
\newblock Springer-Verlag, 1970.

\bibitem{nath75}
P.~Nath, ``On a Coding Theorem Connected with {R{\'{e}}nyi} Entropy,'' {\em
  Information and Control}, vol.~29, pp.~234--242, Nov.~1975.

\bibitem{parker80}
D.S.~Parker, Jr., ``Conditions for Optimality of the {Huffman} Algorithm,''
  {\em SIAM Journal on Computing}, vol.~9, pp.~470--489, Aug.~1980.

\bibitem{Baer2006}
M.B.~Baer, ``A General Framework for Codes Involving Redundancy Minimization,''
  {\em IEEE Transactions on Information Theory}, vol.~IT-52, pp.~344--349, Jan.~2006.

\bibitem{hu-kleitman-tamaki79}
T.C.~Hu, D.J.~Kleitman, and J.K.~Tamaki, ``Binary Trees Optimum Under
  Various Criteria,'' {\em SIAM Journal on Applied Mathematics}, vol.~37,
  pp.~246--256, Apr.~1979.

\bibitem{humblet81}
P.A.~Humblet, ``Generalization of {Huffman} Coding to Minimize the Probability of Buffer Overflow,'' {\em IEEE Transactions on Information Theory}, vol.~IT-27,  pp.~230--237, Mar.~1981.

\bibitem{shtarkov87}
Y. M.~Shtarkov, ``Universal Sequential Coding of Single Messages,'' {\em
  Problemy Peredachi Informatsii}, vol.~23, pp.~175--186, July-Sept.~1987.

\end{thebibliography}

\end{document}